Centre de Physique Théorique* - CNRS - Luminy, Case 907
F-13288 Marseille Cedex 9 - France

# EXISTENCE AND STABILITY OF STEADY FRONTS IN BISTABLE CML

Bastien FERNANDEZ[1]

## Abstract

We prove the existence and we study the stability of the kink-like fixed points in a simple Coupled Map Lattice for which the local dynamics has two stable fixed points. The condition for the existence allows us to define a critical value of the coupling parameter where a (multi) generalized saddle-node bifurcation occurs and destroys these solutions. An extension of the results to other CML's in the same class is also displayed. Finally, we emphasize the property of spatial chaos for small coupling.

The dynamics of localized structures is known to be one of the most relevant features of the extended dynamical systems. These particular solutions usually manifest themselves as solitons, interfaces, fronts, kinks or domain walls separating two regions of the space where the time evolution is homogeneous (or at least regular), namely the domains. The kink propagation in space is often invoked as a destabilization factor for the stable domains and is therefore designed to be the origin of disorder in the underlying system. With the dynamics of spatial wavelenghts, the kink dynamics are thus the main components to be analyzed in the framework of spatio-temporal intermittency [1, 2].

Various models for the dynamics of large systems have been proposed [3]. Most of them consist of Partial Differential Equations (PDE's) such as the Ginzburg-Landau or the Swift-Hohenberg equation. In this framework the problem of the dynamics of fronts is now well-understood in the cases where one domain is stable whereas the other is unstable [4, 5]. Furthermore, models of Ordinary Differential Equations (ODE's) coupled via a discrete laplacian were also introduced as supports for discrete space dynamics (e.g. the Nonlinear Schrödinger Equation). With these systems, the dynamics of interfaces was analytically investigated as a space discretization problem, and hence with the assumption of being governed by PDE's, and the solution exhibit a good agreement with the experimental and numerical data [6, 7].

In this article, we propose an alternate description of the kink dynamics in a (one-dimensional) space-time discrete dynamical system with a continuous state, namely the Coupled Map Lattice (CML) [8, 9]. The simplest kinks, that is to say the fronts in bistable systems, are studied. The numerical simulations reveal the so-called "propagation failure". For a non-symmetric local dynamics, the fronts are stationary solutions until a particular value of the coupling strenght is reached [10]. Above this value, the fronts propagate in the lattice with a traveling wave-like behaviour. The same behaviour is observed in the front (between two stable domains) solutions of PDE's. However the interfaces in these models are moving for all the values of the coupling strength. The difference between both the models, the effect of pinning, can then be assigned to a space discretness. This effect is also reminicent of various phenomena in condensed matter physics such as the Peierls-Nabarro barrier in the Frenkel-Kontorova model of dislocations [11].

The main goal of this paper is to prove the existence of the steady front solutions in a simple CML, until a particular value of the diffusion coefficient is reached, where a (multi) generalized saddle-node bifurcation occurs. In condensed matter physics, the effect of pinning is explained using a two dimensional area-preserving map which represents the action of a dynamical system in space. Following the same idea, we show how it is possible to contruct explicitly the kink solution. The properties of the computed solutions are then examined in order to describe the instabilty which is at the origin of the front propagation. These results are extended to another local map and a numerical investigation is proposed in the case of a continuous nonlinear map. Finally, we focus our attention on the other types of fixed points in these systems.

# 1  Definitions

The "physical space" of the CML under consideration is chosen to be the infinite one-dimensional lattice $\mathbb{Z}$. The phase space is the direct product $M = [0,1]^{\mathbb{Z}}$. A point $x \in M$ is written $x = (x_i)_{i \in \mathbb{Z}}$. We will give below a norm on this phase space in order to give it the structure of a closed subspace of a Banach space.

The CML is a one-parameter family of mappings:

$$F_\epsilon : \quad M \longrightarrow M$$
$$x^t \longmapsto x^{t+1}$$

where $x^t$ denotes the state of the system at time $t$. The model is to be representative of the simplest reaction-diffusion systems, that is to say when the spatial interaction is just the usual (discrete) laplacian operator. The new state at time $t+1$ is then given by the following convex linear combination:

$$x_i^{t+1} = (F_\epsilon x^t)_i = (1-\epsilon) f(x_i^t) + \frac{\epsilon}{2} \Big( f(x_{i-1}^t) + f(x_{i+1}^t) \Big) \quad \forall i \in \mathbb{Z}$$

The parameter $\epsilon \in [0,1]$ is the diffusion coefficient.

The (nonlinear) local map is taken to be the simplest map of the interval that is bistable. To be precise, we introduce a multi-parameter family of piecewise linear mappings (Fig. 1):

$$f_{\{\mu\}} : \quad [0,1] \longrightarrow [0,1]$$
$$x \longmapsto f(x)$$

such that $\{\mu\} = \{a, \alpha, \beta, c, \epsilon\}$ and

$$f(x) = \begin{cases} ax + \alpha & \text{if} \quad 0 \le x < c \\ f(c) & \text{if} \quad x = c \\ ax + \beta & \text{if} \quad c < x \le 1 \end{cases} \quad (1)$$

where

$$f(c) = \min \{ \frac{2ac + (\alpha + \beta)(1 - \sqrt{1 + \frac{2a\epsilon}{1-a}})}{2(1 - \sqrt{(1-a)(1-a+2a\epsilon)})}, ac + \beta \}$$

depends on $\epsilon$. $f(c)$ is defined in a way such that, as will be shown later, there always exists an unstable kink-like fixed point when the corresponding stable one exists. The parameters $a, \alpha, \beta$ and $c$ obey the following inequalities:

$$\begin{cases} 0 < a < 1 \\ 0 < \alpha < \beta < 1 \\ \alpha < c(1-a) < \beta \end{cases}$$

These ensure the existence of two stable fixed points for $f$:

$$X^1 = \frac{\alpha}{1-a} \quad \text{and} \quad X^2 = \frac{\beta}{1-a}$$

For the sake of simplicity, we define $f$ so that these fixed points are the only attractors. The choice of this particular map is motivated by its simplicity which allows us to handle analytically some aspects of the CML dynamics. Notice that we are not dealing with a simpler CML where the local map is piecewise constant, that is to say $a = 0$, because in that case, the model reduces to a Cellular Automata model, that is to say, to a finite set of states.

It is also possible to compute the kink-like fixed points for a more general situation where the local dynamics is continuous:

$$g(x) = \begin{cases} ax + \alpha & \text{if} \quad 0 \leq x < c_1 \\ cx + \gamma & \text{if} \quad c_1 \leq x \leq c_2 \\ ax + \beta & \text{if} \quad c_2 < x \leq 1 \end{cases} \quad (2)$$

with the conditions $0 < a < 1$ and $c > 1$, which ensure the existence and give the stability of the three fixed points (the constants are also fixed so that $g$ is continuous):

$$\frac{\alpha}{1-a} < \frac{\gamma}{1-c} < \frac{\beta}{1-a}$$

The results on the existence and the description of the kink solutions for the map $g$ are similar to that obtained below for the map $f$ and we give the final results below. However the failure of continuity of $f$ may prevent the extension of some results presented here, such as some of the stated properties of the trajectories in the phase space.

In the following, the parameters $a, \alpha, \beta$ and $c$ in (1) are fixed and the study consists in varying the diffusive coefficient $\epsilon$ in order to describe the symmetry breaking in the set of kink solutions, that is to say, the front bifurcation that generally develops in this particular bistable dynamical system.

The local map $f$ is a non-differentiable mapping; more precisely, since the loss of differentiability occurs at $c$, this point plays a central role in this transition. The CML mapping is then non-differentiable (when the state vector has a component equal to $c$) and it is not possible to apply the bifurcation theorems in this case. However, we are able to construct the kink solution using the method of transfer matrices and to sketch the mechanism for the bifurcation that leads to the propagating front structures.

## 2 The kink solutions

At first, we define a kink to be an orbit $\{x^t\}_{t \in \mathbb{N}}$ of the CML with the properties:

$$x_i^t \leq x_{i+1}^t \quad \forall i$$

and

$$\lim_{i \to -\infty} x_i^t = X^1, \quad \lim_{i \to +\infty} x_i^t = X^2$$

The set of kink $K$ is an invariant set: $F_\epsilon(K) \subset K$. This comes from the fact that the local map (1) is an increasing function on $[0, 1]$ and that $X^1$ and $X^2$ are fixed points.

We now consider the set $S(\epsilon)$ of the steady kink solutions:

$$S(\epsilon) = S'(\epsilon) \cup \{x^-, x^+\}$$

where

$$S'(\epsilon) = \{x^t \in K | \ x^t = x \ \forall t\}$$

and $x^-$ (resp. $x^+$) is the homogeneous solution defined by

$$x_i^- = X^2 \ \forall i \ (\text{resp. } x_i^+ = X^1 \ \forall i)$$

Notice that, by definition, the kink solutions $x \in S(\epsilon)$ obey the fixed point equation $F_\epsilon x = x$.

Of course, the present study is also valid for the anti-kink orbits for which one has to consider instead of $K$ the set:

$$AK = \{x^t \in M | \ x_i^t > x_{i+1}^t \ \forall i, \ \lim_{i \to -\infty} x_i = X^2, \ \lim_{i \to +\infty} x_i = X^1\}$$

$S'(\epsilon)$ is decomposed into the following disjoint subsets which we consider separately:

$$S'(\epsilon) = S'_s(\epsilon) \cup S'_u(\epsilon)$$

where

$$S'_s(\epsilon) = \{x \in S'(\epsilon) | \ \forall i \ x_i \neq c\}$$

and

$$S'_u(\epsilon) = \{x \in S'(\epsilon) | \ \exists j \ x_j = c\}$$

Let $T$ be the space translation operator:

$$T: \ M \longrightarrow M$$
$$x \longmapsto Tx$$

where $(Tx)_i = x_{i+1} \ \forall i$.

One can check that $F_\epsilon$ and $T$ commute. Consequently, the subsets $S'_s(\epsilon)$ and $S'_u(\epsilon)$ are (globally) invariant under the action of $T$. We shall see that each of these subsets is entirely determined by any given element, i.e. each is the orbit under $T$ of a single fixed point. If one also notes that the homogeneous fixed points are (pointwise) invariant under the space translations, one can deduce the (global) invariance of $S(\epsilon)$ under the actions of $T$.

Let $j \in \mathbb{Z}$. For the sake of simplicity, we denote by $x_s^j \in S'_s(\epsilon)$ and $x_u^j \in S'_u(\epsilon)$ the particular kink solutions with the properties:

$$\begin{cases} x_{s,i}^j < c & \forall i < j \\ x_{s,i}^j > c & \forall i \geq j \end{cases} \qquad (3a)$$

and

$$\begin{cases} x_{u,i}^j < c & \forall i < j \\ x_{u,j}^j = c \\ x_{u,i}^j > c & \forall i > j \end{cases} \qquad (3b)$$

According to these definitions, the set $S'(\epsilon)$ is explicitly given by:

$$S'(\epsilon) = \bigcup_{j \in \mathbb{Z}} \{x_s^j, x_u^j\}$$

since we shall prove that, for each $j$, there is a unique $x_s^j$ and $x_u^j$ in $S'(\epsilon)$.

We now describe the computation of $x_s^j$ using a two dimensional area-preserving map. First, we intoduce the deviation $y = (y_i)_{i \in \mathbb{Z}}$ from the (local map) fixed points:

$$y_i = \begin{cases} x_{s,i}^j - \dfrac{\alpha}{1-a} & \forall i < j \\ \dfrac{\beta}{1-a} - x_{s,i}^j & \forall i \geq j \end{cases}$$

The computation of $x_s^j$ components then reduces to the problem of determining the sequences of vectors of the plane that are related by the linear transformations:

$$\begin{pmatrix} y_{i-1} \\ y_i \end{pmatrix} = A_\epsilon \begin{pmatrix} y_i \\ y_{i+1} \end{pmatrix} \quad \forall i < j-1 \tag{4a}$$

and

$$\begin{pmatrix} y_{i+1} \\ y_i \end{pmatrix} = A_\epsilon \begin{pmatrix} y_i \\ y_{i-1} \end{pmatrix} \quad \forall i > j \tag{4b}$$

where $A_\epsilon = \begin{pmatrix} \frac{2}{a\epsilon}(1-a+a\epsilon) & -1 \\ 1 & 0 \end{pmatrix}$ is a $2 \times 2$ hyperbolic matrix.

According to the boundary conditions for the elements of $S'_s(\epsilon)$, the $y$ components must vanish at both $+\infty$ and $-\infty$. This implies that the vectors in the relations (4) must be in the contracting (eigen-) direction of $A_\epsilon$. Therefore the components of $y$ are:

$$y_i = \begin{cases} y_{j-1}(\lambda_-)^{j-i-1} & \forall i < j \\ y_j(\lambda_-)^{i-j} & \forall i \geq j \end{cases}$$

where $\lambda_-$ is the eigenvalue of $A_\epsilon$ that is less than one:

$$\lambda_- = \frac{1 - a + a\epsilon - \sqrt{(1-a)(1-a+2a\epsilon)}}{a\epsilon}$$

The computation of the constants $y_{j-1}$ and $y_j$ is performed by writing the affine transformations which corresponds to the connection between the sites above $c$ and those below $c$:

$$\begin{pmatrix} y_j \\ y_{j-1} \end{pmatrix} = \begin{pmatrix} -\frac{2}{a\epsilon}(1-a+a\epsilon) & 1 \\ 1 & 0 \end{pmatrix} \begin{pmatrix} y_{j-1} \\ y_{j-2} \end{pmatrix} + \begin{pmatrix} \frac{\beta-\alpha}{a(1-a)} \\ 0 \end{pmatrix} \tag{5a}$$

$$\begin{pmatrix} y_{j+1} \\ y_j \end{pmatrix} = \begin{pmatrix} \frac{2}{a\epsilon}(1-a+a\epsilon) & 1 \\ 1 & 0 \end{pmatrix} \begin{pmatrix} y_j \\ y_{j-1} \end{pmatrix} - \begin{pmatrix} \frac{\beta-\alpha}{a(1-a)} \\ 0 \end{pmatrix} \tag{5b}$$

The composition of the relation (5) leads to a new affine transformation between the initial vectors of the maps (4). This new map formally reads:

$$\begin{pmatrix} y_{j+1} \\ y_j \end{pmatrix} = \mathcal{A} \begin{pmatrix} y_{j-1} \\ y_{j-2} \end{pmatrix} + T$$

The matrix $\mathcal{A}$ and the vector $T$ are obviously deduced from those of relation (5). In order to ensure the existence and consequently the uniqueness of the solution, we check using the properties:

$$\begin{pmatrix} y_{j+1} \\ y_j \end{pmatrix} = y_j \begin{pmatrix} \lambda_- \\ 1 \end{pmatrix} \text{ and } \begin{pmatrix} y_{j-1} \\ y_{j-2} \end{pmatrix} = y_{j-1} \begin{pmatrix} 1 \\ \lambda_- \end{pmatrix}$$

that the vectors

$$\begin{pmatrix} y_{j+1} \\ y_j \end{pmatrix} \text{ and } \mathcal{A} \begin{pmatrix} y_{j-1} \\ y_{j-2} \end{pmatrix}$$

are linearly independent. This calculation leads to the simple condition $a < 1$. Therefore the existence and uniqueness of the kink solution is given by the stability of the local map fixed points. The computation effectively reduces to the resolution of a system of two linear equations in $y_{j-1}$ and $y_j$ and the components of $x_s^j$ finally read:

$$x_{s,i}^j = \begin{cases} \dfrac{\alpha}{1-a} + \dfrac{\beta - \alpha}{a(1-a)(1+\lambda_-)}(\lambda_-)^{j-i} & \forall i < j \\ \dfrac{\beta}{1-a} - \dfrac{\beta - \alpha}{a(1-a)(1+\lambda_-)}(\lambda_-)^{i-j+1} & \forall i \geq j \end{cases} \quad (6)$$

However, some restrictions on this solution may be imposed by the conditions (3a). Actually it will be shown below that these limitations are crucial as they give the bifurcation point, since (3a) and (6) are compatible only if $\epsilon$ is smaller than a critical value defined in section 4.

The same method is applied in order to compute the components of $x_u^j$. Also in this case the determination of the constants $y_{j-1}$ and $y_{j+1}$ gives an equation similar to the system (5). Thanks to the particular definition of $f(c)$, the condition for the existence and uniqueness of the solution is also $a < 1$ and $x_u^j$ reads:

$$x_{u,i}^j = \begin{cases} \dfrac{\alpha}{1-a} + \dfrac{1}{a}(f(c) - \dfrac{\alpha}{1-a})(\lambda_-)^{j-i} & \forall i < j \\ c & i = j \\ \dfrac{\beta}{1-a} - \dfrac{1}{a}(\dfrac{\beta}{1-a} - f(c))(\lambda_-)^{i-j} & \forall i > j \end{cases} \quad (7)$$

A plot of these solutions is given in Figure 2. Notice that the uniqueness also implies that the system (3b) has a unique solution for $\epsilon < \epsilon_c$, where the critical value $\epsilon_c$ will be specified below in section 4.

For the corresponding anti-kink structures, the elements of $S''(\epsilon)$ can be deduced from those of $S'(\epsilon)$ by simply applying the symmetries $R_s^j$ for the points $x_s^j$ and $R_u^j$ for $x_u^j$ where:

$$(R_s^j x)_{j+i} = x_{j-i-1} \quad \forall i$$

and

$$(R_u^j x)_{j+i} = x_{j-i} \quad \forall i$$

## 3  The stability analysis

The stability of the fixed points $x_s^j$ and $x_u^j$ is now investigated. We show that $x_s^j$ is stable whereas $x_u^j$ is unstable; more precisely the former is a saddle. Here again, due to the translational invariance along the lattice, the following results may, with some caution, be extended to any other element of $S_s'(\epsilon)$ and $S_u'(\epsilon)$. The study is performed by the computation of the perturbation dynamics in a neighborhood of these solutions. This approach naturally leads to the description of the stable manifold of $x_s^j$ and $x_u^j$. Noticing that the central manifold is empty, the unstable manifold may be deduced from the stable one.

The sets of perturbations under consideration are

$$V_s^j = \{P \in \mathbb{R}^{\mathbb{Z}} \mid x = x_s^j + P, x \in M, \ x_i < c \ \forall i < j \text{ and } x_i > c \ \forall i \geq j\}$$

in the stability analysis of $x_s^j$ and

$$V_u^j = \{P \in \mathbb{R}^{\mathbb{Z}} \mid x = x_u^j + P, x \in M, \ x_i < c \ \forall i < j, \ x_j \in [0,1] \text{ and } x_i > c \ \forall i > j\}$$

for the point $x_u^j$. These sets overlap from one fixed point to the other.

We adopt the usual definition of the local stable manifold but we may express it in terms of perturbations:

$$W_{loc}^s(x_*^j) = x_*^j + \{P \in V_*^j \mid \mathcal{F}_\epsilon^t(P) \to 0 \text{ as } t \to +\infty \text{ and } \mathcal{F}_\epsilon^t(P) \in V_*^j \ \forall t \geq 0\}$$

where $*$ stands either for $s$ or $u$ and the perturbation map is defined by:

$$\begin{aligned}\mathcal{F}_\epsilon : \ V_*^j &\longrightarrow \mathbb{R}^{\mathbb{Z}} \\ P &\longmapsto \mathcal{F}_\epsilon(P) = F_\epsilon(x_*^j + P) - x_*^j\end{aligned}$$

Moreover the stable manifold is:

$$W^s(x_*^j) = \bigcup_{t \geq 0} \mathcal{F}_\epsilon^{-t}(W_{loc}^s(x_*^j))$$

This definition is independent of the original neighborhood. The computation of these sets is cumbersome and relatively useless. Indeed, the invariant manifolds in the case of maps are generically sets of isolated points, and thus are not manifolds in the usual sense. We avoid the problem of describing all the trajectories by restricting the initial conditions for which the orbits stay in the neighborhoods $x_s^j + V_s^j$ and $x_u^j + V_u^j$.

In the case of the point $x_s^j$ we obtain $\mathcal{F}_\epsilon(P) = J_\epsilon P$ where $J_\epsilon$ is the tridiagonal (infinite-dimensional) operator:

$$J_\epsilon = \begin{pmatrix} \ddots & \ddots & \ddots & & \ddots & & \ddots \\ & 0 & \frac{\epsilon a}{2} & (1-\epsilon)a & \frac{\epsilon a}{2} & 0 & \\ & & \ddots & \ddots & \ddots & \ddots & \ddots \end{pmatrix}$$

For $x_u^j$ the dynamics is more subtle as it contains the local map discontinuity:

$$\forall P \in V_u^j \quad \mathcal{F}_\epsilon(P) = \begin{cases} J_\epsilon P + \eta_- & \text{if } P_j < 0 \\ J_\epsilon P & \text{if } P_j = 0 \\ J_\epsilon P + \eta_+ & \text{if } P_j > 0 \end{cases}$$

where $\eta_-$ and $\eta_+$ are the vectors of $\mathbb{R}^{\mathbb{Z}}$ whose only non-vanishing components are the following:

$$\eta_{-,i} = \begin{cases} \frac{\epsilon}{2}(ac + \alpha - f(c)) & i = j \pm 1 \\ (1-\epsilon)(ac + \alpha - f(c)) & i = j \end{cases}$$

and

$$\eta_{+,i} = \begin{cases} \frac{\epsilon}{2}(ac + \beta - f(c)) & i = j \pm 1 \\ (1-\epsilon)(ac + \beta - f(c)) & i = j \end{cases}$$

The stability of $x_s^j$ is given by the following:

**Proposition (3.1).** $W_{loc}^s(x_s^j) = x_s^j + V_s^j$

This assertion implies that $W_{loc}^u(x_s^j)$ is empty, hence $x_s^j$ is stable, i.e. it is a node. This is the observed solution in numerical simulations.

*Proof.* Here we use the notation $V_s^j$ for $x_s^j + V_s^j$.
- By construction, one has: $W_{loc}^s(x_s^j) \subset V_s^j$.
- We endow $\mathbb{R}^{\mathbb{Z}}$ with the inner product [12]:

$$\langle x, y \rangle_q = \sum_{i \in \mathbb{Z}} \frac{x_i y_i}{q^{|i|}} \text{ , for any } q > 1$$

and the norm $\|.\|_q = \sqrt{\langle ., . \rangle_q}$. We consider the Hilbert space $B_q = \{P \in \mathbb{R}^{\mathbb{Z}} \mid \|P\|_q < \infty\}$ and $\|.\|$ the usual supremum norm for operators. We have

$$\|\mathcal{F}_\epsilon(P)\|_q \leq \|J_\epsilon\|.\|P\|_q, \forall P \in B_q.$$

$J_\epsilon$ is a normal operator, hence we get

$$\|J_\epsilon\| = r(J_\epsilon) = \sup_{\lambda \in \sigma(J_\epsilon)} |\lambda|$$

where $\sigma(J_\epsilon)$ is the spectrum of $J_\epsilon$. Using the method developed in [12], one can deduce that:

$$\sigma(J_\epsilon) \subset \text{Clos}\{\bigcup_{N \geq N_0} \Lambda_N\} \cup \{0\}$$

where
$$\Lambda_N = \{a(1-\epsilon) + a\epsilon \cos\frac{k\pi}{N+1},\ k=1,N\}$$
is the spectrum of the finite-dimensional approximation of $J_\epsilon$, that is to say, the spectrum of the tridiagonal matrix of size $N$.

Therefore $r(J_\epsilon) = |a| = a < 1$. This gives the required statement $\mathcal{F}_\epsilon(P) \in V_s^j\ \forall P \in V_s^j$. The first condition for an element of $V_s^j$ to belong to $W_{loc}^s(x_s^j)$ has been checked. The second one is also valid when writing:

$$\mathcal{F}_\epsilon^t(P) = (J_\epsilon)^t P$$

from which, it clearly follows that $\mathcal{F}_\epsilon^t(P) \to 0$ as $t \to +\infty$. Both of these assertions imply $V_s^j \subset W_{loc}^s(x_s^j)$, which ends the proof. □

We consider the useful properties for the decomposition of $V_*^j$:

**Definition (3.2).** $P \in V_*^j$ is symmetric (resp. skew-symmetric) iff $P_{j+i} = P_{j-i}\ \forall i$ (resp. $P_{j+i} = -P_{j-i}\ \forall i$). The symmetric (resp. skew-symmetric) vectors are denoted $P_s$ (resp. $P_a$).

This definition is motivated by the conservation of some symmetries under the action of $J_\epsilon$; clearly $J_\epsilon.P_s$ is symmetric and $J_\epsilon.P_a$ is skew-symmetric.

Let $V_a^j = \{P \in V_u^j \mid P = P_a\}$, the subset of skew-symmetric perturbations. The stability of $x_u^j$ is given by the following:

**Proposition (3.3).** $W_{loc}^s(x_u^j) = x_u^j + V_a^j$

*Proof.* - By induction. One has $\forall P \in V_a^j\ P_j = 0$. Then $\mathcal{F}_\epsilon(P) = J_\epsilon P$ which is known to have the required properties of being in the local stable manifold.
Moreover, suppose that $\mathcal{F}_\epsilon^t(P) \in V_a^j$; then

$$\|\mathcal{F}_\epsilon^{t+1}(P)\|_q = \|\mathcal{F}_\epsilon(\mathcal{F}_\epsilon(P))\|_q \leq \|J_\epsilon\|\|\mathcal{F}_\epsilon(P)\|_q$$

from which we deduce $V_a^j \subset W_{loc}^s(x_u^j)$ where we discard $x_u^j$ in the notation.
The proof that $W_{loc}^s(x_u^j) \subset V_a^j$ is similar to the previous one. We show that in order to be (always) decreasing and asymptotically vanishing, a perturbation must have a vanishing $j^{th}$ component after each iteration. This condition implies the result. □

Hence we have shown the point $x_u^j$ to be a saddle. A 3-dimensional schematic reprentation of the phase portrait of these two fixed points is displayed in Figure 3. The connection to the other points may not be so easy be cause the symmetric and skew-symmetric axes vary from one stable (or unstable) fixed point to the other.

Finally we have the following:

**Proposition (3.4).** $\forall P \in V_*^j$ such that $\forall t\ \mathcal{F}_\epsilon^t(P) \in V_*^j,\ \lim_{t \to +\infty}\|\mathcal{F}_\epsilon^t(P)\|_q \leq D$
where $D = \max\{\|x_s^j - x_u^j\|_q, \|x_s^{j+1} - x_u^j\|_q\}$

This statement ensures that all suitable orbits evolve towards a point in $S'(\epsilon)$, and thus cannot reach the homogeneous state. (We have not investigated the cases where the orbit leave the neighborhood under consideration, but we conjecture that it will be trapped in a neighborhood of another fixed solution in $S'(\epsilon)$ and stay inside forever.)

*Proof.* From the proof of Proposition (3.1), the case $P \in V_s^j$ obeys the statement, since for such perturbations, the dynamics is simply given by the product with the contracting matrix $J_\epsilon$. In this situation, the limit in norm is zero and the asymptotic state is $x_s^j$.

The case $P \in V_d^j$ also follows from Proposition (3.3) by the same argument and the final state is $x_u^j$.

Assume that $P \in V_u^j$ is such that $P_i > 0 \; \forall i$ (resp. $P_i < 0 \; \forall i$). Then for $\|J_\epsilon\| < 1$, the asymptotic state for such a perturbation is given by:

$$\lim_{t \to +\infty} \mathcal{F}_\epsilon^t(P) = \sum_{t=0}^{\infty} J_\epsilon^t \eta_+ = (Id - J_\epsilon)^{-1} \eta_+ = x_s^j - x_u^j$$

(resp.
$$\lim_{t \to +\infty} \mathcal{F}_\epsilon^t(P) = (Id - J_\epsilon)^{-1} \eta_- = x_s^{j+1} - x_u^j )$$

$Id$ means for the identity operator in $V_u^j$. Consequently, the norm limit is one of the values of $D$ depending on the sign of the $P$ components.

Now, for any $P \in V_u^j$, one has in a componentwise sense:

$$J_\epsilon^t + \sum_{k=0}^{t-1} J_\epsilon^k \eta_- \leq \mathcal{F}_\epsilon^t(P) \leq J_\epsilon^t + \sum_{k=0}^{t-1} J_\epsilon^k \eta_+ \; \forall t > 0$$

This inequality implies the asymptotic boundedness. □

## 4 The conditions for the existence of the kink-like fixed points and the generalized saddle-node bifurcation

The kink-like fixed points have been computed assuming the conditions (3). However these assumptions have to be checked afterwards as the expressions (6) and (7) for the components of $x_s^j$ and $x_u^j$ mainly depend on $\epsilon$. Some properties of the components (6) will allow us to claim a criterion for the existence of the fixed interfaces in our CML. One can check that the expressions (6) obey

**Proposition (4.1).** - $\forall i < j$ (respectively $\forall i \geq j$) the components of $x_s^j$ are increasing (resp. decreasing) functions of the coupling strength $\epsilon$.

- Let $K_\alpha \equiv \frac{c(1-a)-\alpha}{\beta-\alpha}$, $K_\beta \equiv \frac{\beta-c(1-a)}{\beta-\alpha}$ and $\epsilon_{\alpha,\beta} \equiv \frac{2(1-a)K_{\alpha,\beta}(1-aK_{\alpha,\beta})}{(1-2aK_{\alpha,\beta})^2}$; the following is true:

$$\forall \epsilon > \epsilon_\alpha \quad x_{s,j-1}^j > c$$
$$\forall \epsilon > \epsilon_\beta \quad x_{s,j}^j < c$$

The proof is accomplished with simple calculations.

This proposition implies that the fixed points $x_s^j$ no longer exist for $\epsilon > \epsilon_c \equiv \min\{\epsilon_\alpha, \epsilon_\beta\}$. Moreover the image of $c$ has been constructed in such a way that the points $x_u^j$ also do not exist any more when $\epsilon > \epsilon_c$. This is because the problem of the transfer matrices for the saddle points has no solution for this range of diffusive coefficient as $f(c) = ac + \beta$.

In other words, we have described a (multi) generalized saddle-node bifurcation that occurs for all the kink-like fixed points in our bistable CML. This bifurcation can be viewed as a transition from a global translational symmetry invariance in the set of fixed interfaces

$$S(\epsilon) = \bigcup_{j \in \mathbb{Z}} \{x_s^j, x_u^j\} \cup \{x^-, x^+\} \ \forall \ 0 \leq \epsilon \leq \epsilon_c$$

to a pointwise translational symmetry of

$$S(\epsilon) = \{x^-, x^+\} \ \forall \epsilon_c < \epsilon \leq 1$$

The resulting attractors for a kink-like initial condition may be one of the homogeneous solutions when $\epsilon > \epsilon_c$. Indeed the analysis of the perturbation dynamics near a fixed point $x^j \equiv x_s^j = x_u^j$ at $\epsilon = \epsilon_c$ may give an insight into this property.

Let $c > (X^1 + X^2)/2$; then $\epsilon_c = \epsilon_\beta$. Note that the case of equality is the symmetric case where the fixed fronts always exist (that is for any $\epsilon \in [0,1]$), and that the case $c < (X^1 + X^2)/2$ is achieved in the same way. The dynamics for a perturbation of the fixed point $x^j$ reads:

$$\mathcal{F}_{\epsilon_c}(P) = \begin{cases} J_{\epsilon_c} P & \text{if } P_j \geq 0 \\ J_{\epsilon_c} P + \nu_-^j & \text{if } P_j < 0 \end{cases}$$

where $\nu_-^j$ is the vector $\eta_-$ computed at $\epsilon_c$:

$$(\nu_-^j)_i = \begin{cases} 0 \text{ if } i < j-1 \text{ or } i > j+1 \\ \frac{\epsilon_c}{2}(\alpha - \beta) & i = j \pm 1 \\ (1 - \epsilon_c)(\alpha - \beta) & i = j \end{cases}$$

Every perturbation with positive components is damped and the asymptotic state is $x^j$. However, any perturbation with negative components is not damped and as time evolves, it approaches the value (see the proof of (3.4)):

$$\lim_{t \to +\infty} \mathcal{F}_{\epsilon_c}^t(P) = x^{j+1} - x^j$$

The asymptotic state of the system is $x^{j+1}$ in this case. Therefore any kink-like initial condition, that is to say any initial condition in the basin of attraction of one of the $x^j$, evolves towards the "right" (from the lattice point of view) and reaches one of the fixed points asymptotically. By contrast, any anti-kink like initial condition may propagate to the "left", as can be seen from a similar perturbation analysis of points in $S''(\epsilon)$ at $\epsilon = \epsilon_c$. Hence, according to the sign of the quantity $c - \frac{X^1 + X^2}{2}$, one might decide on the direction of the front and the anti-front propagation for the coupling above the critical value.

# 5 The transition for continuous local maps

In this section, we describe the steady-propagating front transition for continuous local maps. The first situation deals with the map $g$ (defined in (2)), then we consider numerically the case of a differentiable mapping.

If the local map is chosen to be the map $g$, the same analysis as above can be done, that is to say, the calculation of the points $x_s^j$ and $x_u^j$, the analysis of stability and the bifurcation. In this case, the $x_s^j$ components are also given by the system (6) whereas we only consider the unstable solution with one component in the interval $]c_1, c_2[$. One can check that $x_u^j$ exists and is unique. The condition for the existence of the steady front is also $\epsilon < \epsilon_c \equiv \min\{\epsilon_\alpha, \epsilon_\beta\}$ where $\epsilon_\alpha$ and $\epsilon_\beta$ are defined as in (4.1) but with different values of $K_\alpha$ and $K_\beta$:

$$K_\alpha = \frac{c_1(1-a) - \alpha}{\beta - \alpha} \text{ and } K_\beta = \frac{\beta - c_2(1-a)}{\beta - \alpha}$$

The stability analysis of $x_s^j$ also implies that it is a stable point. The investigation of the perturbation dynamics is not so simple for $x_u^j$ but again it is possible to show that it is a saddle. This result is confirmed by the numerical computation of the associated linear dynamics spectrum. Hence, the CML dynamics also reveals a generalized saddle-node bifurcation in this case.

One step further in the complexity of the local dynamics is to examine a differentiable bistable map; a model that is closer to a more realistic situation. Here we have chosen the (non-symmetric) cubic map $h_{(\mu,c)}(x) = c + \mu x(1 - x^2)$. For suitable values of $\mu$ and $c$, $h_{(\mu,c)}$ is also bistable. In this context, we have no idea how to compute explicitly the components of the fixed point. Indeed, the method of transfer matrices is no longer appropriate because the relation between the neighbours are quadratic.

However, due to the bistable feature, the CML with the map $h_{(\mu,c)}$ may reveal the same bifurcation as in the former cases. In order to check this claim, we have computed numerically the spectrum of the Jacobian associated with the kink fixed point. The result is presented in Figure 4 where it clearly appears that the greatest Jacobian eigenvalue occurs at one for $\epsilon = 0.1193$, the mark of a saddle-node bifurcation in differentiable cases. This value of the coupling exactly corresponds to the value at which the front propagates in the lattice, as can be seen from the simulations. Notice the interesting and somewhat unexpected result (see Figure 4) that the spectrum shows an isolated eigenvalue that crosses the unit circle and is isolated from the remainder of the spectrum by an uniform gap.

# 6 The other fixed points

In this section, we give an insight into the other types of steady solutions inherent to the CML under consideration, that is to say with the local map (1). The fixed point equation is expressed in the following terms:

$$G(\epsilon, x) = 0 \qquad (8)$$

with $G(\epsilon, x) = F_\epsilon x - x$, and we by denote $\mathcal{S}(\epsilon)$ the set of solutions depending on the coupling strength. We endow the extended phase space $I\!R^{\mathbb{Z}}$ with the usual norm:

$$\|x\|_\infty = \sup_{i \in \mathbb{Z}} |x_i|$$

and from now on we consider the Banach space $B_\infty = \{x \in I\!R^{\mathbb{Z}} \mid \|x\|_\infty < \infty\}$.
For $\epsilon = 0$ the dynamics consists of a set of uncoupled maps $(F_0 x^t)_i = f(x_i^t)$; this yields:

$$\mathcal{S}(0) = \{x \in [0,1]^{\mathbb{Z}} \mid \forall i \ x_i = X^1 \text{or} X^2\}$$

which means that the system has the property of spatial chaos [13].

The continuation of the fixed points into the coupled case is guaranteed by the application of the Implicit Function Theorem to equation (8) at each point $x_0$ of $\mathcal{S}(0)$ [14] [15]. We describe now the conditions for the use of this theorem and its consequence.
Let $U(0, x_0) \in (I\!R, |.|) \times B_\infty$ be an open neighborhood of $(0, x_0)$ such that:

(i) The (infinite) jacobian $DG(0, x_0)$ exists as a Frechet derivative on $U(0, x_0)$ and is invertible.

(ii) $G$ and $DG$ are continuous at $(0, x_0)$.

The conclusion is then that there exists a number $\delta$ such that for every $\epsilon$ satisfying $|\epsilon| < \delta$ there is exactly one $x(\epsilon)$ for which $G(\epsilon, x(\epsilon)) = 0$. Note that, thanks to the linearity of $f$, the theorem also gives an exact bound on $x(\epsilon)$. Furthermore, as $G(\epsilon, x)$ is continuous in a neighborhood of $(0, x_0)$, $x(\epsilon)$ is continuous in a neighborhood of 0.
Here we choose

$$U(0, x_0) = ]0, \delta[ \times \prod_{i \in \mathbb{Z}} I_i$$

where

$$I_i = \begin{cases} ]0, c[ & \text{if } (x_0)_i = X^1 \\ ]c, 1[ & \text{if } (x_0)_i = X^2 \end{cases}$$

The main condition for the continuation of $x_0$ is (i), thus the fixed points may exist as far as $F_\epsilon$ is differentiable. This condition fails when (at least) one component of $x(\epsilon)$ is $c$. Hence we may obtain a condition for the existence of any fixed point similar to the one computed for the kink solution (Proposition (4.1)). The bound $\delta$ depends on the particular point $x_0$ but it is possible to obtain a uniform bound for any solution [14]. Moreover not only is $x(\epsilon)$ unique, but since $F_\epsilon$ is contracting on $U(0, x_0)$ there is no other possible fixed solution in this set. Note also that the Implicit Function Theorem can also be applied in the case of a differentiable local map. In such a case, $G(\epsilon, x)$ is always differentiable and the fixed points exist as long as $DG(\epsilon, x(\epsilon))$ is invertible, that is to say, until the spectrum of $DF_\epsilon$ lies entirely within the unit circle.

Finally, as in the case of the front, the critical values of $\epsilon$ (for which the solution disappears) are given for two examples of the CML defined with the map (1). This is done by generalizing the transfer matrix technique and by checking afterwards the conditions for the existence of the solution. Here we suppose again for the sake of definiteness that $c > (X^1 + X^2)/2$ (the opposite case can be handled in a similar manner). For the one-point domain solution which is defined by

$$\exists j \text{ such that } x_j > c \text{ and } \forall i \neq j \ x_i < c$$

the critical value is

$$\epsilon'_\beta = \frac{(1-a)K_\beta(2-aK_\beta)}{2(1-aK_\beta)^2}$$

where $K_\beta$ is given in (4.1). This solution is (numerically) the less stable fixed point in the structural sense, that is to say, the first solution to disappear when one increases $\epsilon$ from 0. For the (spatial) 2-periodic point

$$\forall i \ x_{2i} > c \text{ and } x_{2i+1} < c$$

we have found

$$\epsilon''_\beta = \frac{(1-a)K_\beta}{1-2aK_\beta}$$

and we obtain the following ordering of the critical values

$$\epsilon'_\beta < \epsilon''_\beta < \epsilon_\beta$$

from which we conjecture the kink solution to have the largest transition value.

### Acknoweldgements
I acknowledge R. Lima for relevant comments and A. Lambert for his help in the numerical part of this work.

# Figure Captions

**Figure 1:** The local map $f$. The plot is for the situation where $c > (X^1+X^2)/2$ and $\epsilon < \epsilon_\beta$ (see section 4).

**Figure 2:** The stable $x_s^j$ and the unstable $x_u^j$ fixed points for $\epsilon = 0.3$. The parameters for $f$ are $a = 0.4$, $\alpha = 0.1$, $\beta = 0.5$ and $c = 0.65$. The horizontal lines stand respectively for $X^1$, $c$ and $X^2$.

**Figure 3:** Schematic 3-dimensional representation of a phase space region. The solid lines stand for the stable directions of the (local) stable manifold. The dot-dashed lines represent the unstable local manifold.

**Figure 4:** Spectrum of the kink-like fixed point Jacobian (Imaginary part vs. Real part). The parameters are equal to: $\mu = 1.3$, $c = 0.02$ and $\epsilon = 0.1193$.

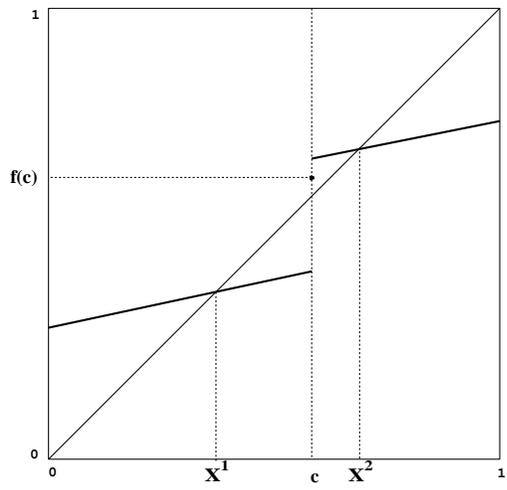

Figure 1:

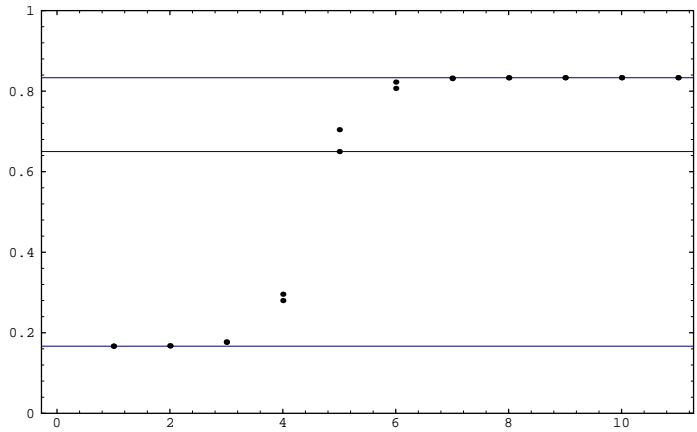

Figure 2:

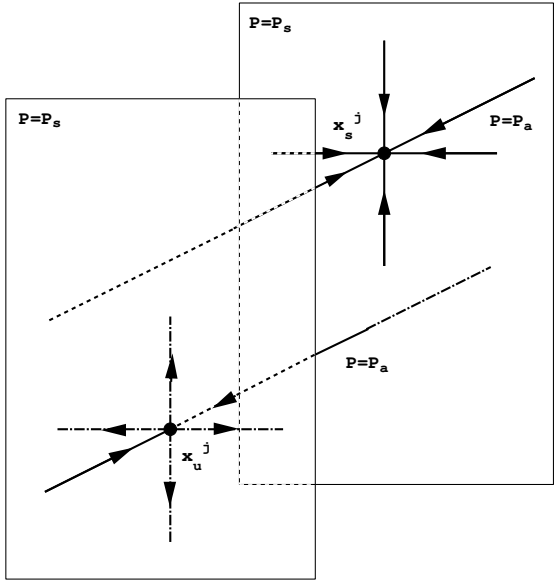

Figure 3:

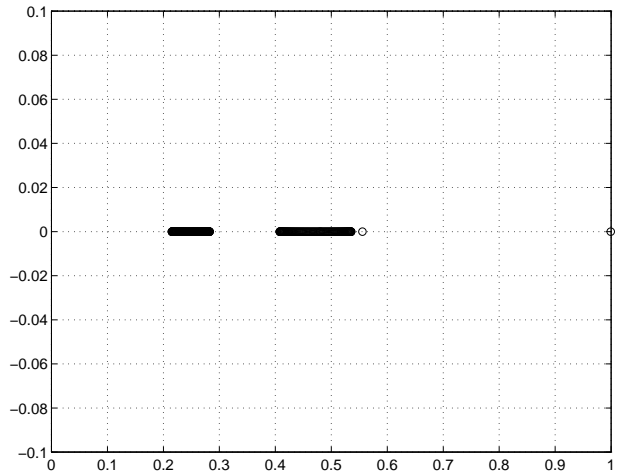

Figure 4: